\documentclass[preprint,secnumarabic,superscriptaddress,amssymb, nobibnotes, aps, prb]{revtex4-1}
\usepackage{graphicx}

\begin{document}

\title{Tuning Photoinduced Terahertz Conductivity in Monolayer Graphene: Optical Pump Terahertz Probe Spectroscopy}
\author{Srabani Kar}
\affiliation{Department of Physics, Indian Institute of Science, Bangalore 560 012, India}
\affiliation{Center for Ultrafast Laser Applications, Indian Institute of Science, Bangalore 560 012, India}
\author{Dipti R. Mohapatra}
\affiliation{Department of Physics, Indian Institute of Science, Bangalore 560 012, India}
\author{Eric Freysz}
\affiliation{University of Bordeaux, LOMA, UMR CNRS 5798, 351, Cours de la liberation, 33405 Talence cedex, France}
\author{A. K. Sood}
\email[corresponding author:]{asood@physics.iisc.ernet.in}
\affiliation{Department of Physics, Indian Institute of Science, Bangalore 560 012, India}
\affiliation{Center for Ultrafast Laser Applications, Indian Institute of Science, Bangalore 560 012, India}

\begin{abstract}
 Optical pump-terahertz probe differential transmission measurements of as-prepared single layer graphene (AG) (unintentionally hole doped with Fermi energy  $E_F$ at $\sim$180 meV), nitrogen doping compensated graphene (NDG) with $E_F$ $\sim$10 meV and thermally annealed doped graphene (TAG) are examined quantitatively to understand the opposite  signs  of photo-induced dynamic  terahertz conductivity  $\Delta\sigma$. It  is negative  for AG and TAG but positive for NDG. We show that the recently proposed mechanism of multiple generations of secondary hot carriers due to Coulomb interaction of photoexcited carriers with the existing carriers together with the intraband scattering can explain the change of photoinduced conductivity sign and its magnitude. We give a quantitative estimate of $\Delta\sigma$ in terms of controlling parameters - the Fermi energy $E_F$ and momentum relaxation time $\tau$. Further, the cooling of photoexcited carriers is analyzed using super–collision model which involves defect mediated collision of the hot carriers with the acoustic phonons, thus giving an estimate of the deformation potential. 
\end{abstract}
\maketitle

\section{Introduction}
The performance of electronic and optoelectronic devices depends on the material properties such as carrier mobility, energy conversion efficiency from photon to electron-hole pairs, spectral response and equilibration of the photo-generated carriers. The linear band dispersion with zero band gap in monolayer graphene responsible for many fascinating transport phenomena and optical effects makes graphene a desirable material for high speed optoelectronic devices.\cite{Bonaccorso2010,Novoselov2005,Morozov2008,Avouris2010,Geim2007,Weis2012,Bao2012} A key question to answer in optoelectronic applications is the relaxation of the hot carriers in conical energy-momentum space of the monolayer graphene. Ultrafast time resolved pump-probe spectroscopy 
\cite{Wang2010,Choi2009,Dawlaty2008,George2008,Sun2010} and angle resolved photoemission spectroscopy
\cite{Gierz2013} have been shown to be excellent probes of non-equilibrium carrier dynamics in graphene. The terahertz probe pulse following the optical pump examines the intraband scattering dynamics as compared to optical probe which is sensitive to both interband and intraband scattering processes. Following the pump pulse, the photoexcited carriers achieve a quasi-equilibrium Fermi-Dirac distribution characterized by electron temperature $T_e$ mostly by carrier-carrier scattering. The cooling of the carriers can occur through phonon emission as well as through carrier-carrier scattering. The latter involves the transfer of energy of photoexcited carriers to the existing carriers in graphene (making them hot).
\cite{Song201287,Song2011,Tielrooij2013} The cooling via phonon emission involving optical phonons occurs on a time scale of $\sim$100 to 500 fs till the energy of the photoexcited carriers is less than the optical phonon energy ($\sim$200 meV). This is followed by the direct coupling between the carriers and acoustic phonons which can last for tens of ps. However the carrier-acoustic phonon relaxation time is reduced to less than 10 ps when large momentum and large energy acoustic phonons emission occurs mediated by the disorder. This three body (carriers + acoustic phonons + disorder) mediated cooling is termed as super collision cooling of the carriers.
\cite{Graham2013} Several groups have reported optical pump - terahertz probe (OPTP) time domain spectroscopy of epitaxial grown as well as CVD grown graphene showing positive
\cite{Choi2009,Sun2010,Strait2011} as well as negative dynamic conductivity.
\cite{Tielrooij2013,Docherty2012,Jnawali2013,Frenzel2013} The positive dynamic conductivity can be easily understood in terms of intraband scattering of the carriers. Docherty et al.
\cite{Docherty2012} showed that the THz photoconductivity changes from positive in vacuum to negative in nitrogen, air and oxygen environment and proposed stimulated THz emission from photoexcited graphene as the cause of the negative photoconductivity. Several other experimental
\cite{Karasawa2011} and theoretical
\cite{Ryzhii2007,Satou2013} studies have also attributed the negative dynamic conductivity to the amplified stimulated terahertz emission above a threshold pump intensity.
\cite{Otsuji2012} In this context, Gierz et al.
\cite{Gierz2013} have shown experimentally that only within 130 fs after photoexcitation, the Fermi-Dirac distribution for electron and holes are different, suggesting that population inversion and hence stimulated emission is not feasible beyond this time window. Jnawali et al.
\cite{Jnawali2013} have attributed the decrease in photoconductivity to the increase in carrier scattering rate with negligible increase of Drude weight. Tielrooij et al.\cite{Tielrooij2013} have proposed to explain the negative dynamic conductivity via Coulomb interaction governed carrier-carrier scattering where the energy of the photo-excited carriers is transferred to the existing carriers in the Dirac cone, a process termed as secondary hot carrier generation (SHCG). While preparing our manuscript we came across a recent study of tuning the sign of OPTP signal from the SLG using electrostatic top gating.
\cite{Shi2014} Their explanation is as follows: Taking conductivity $\sigma = \mathfrak{D}/ \Gamma$, the dynamic conductivity $\Delta\sigma = (\Delta\mathfrak{D}/{\mathfrak{D}_0}){\sigma_0}  - (\Delta\Gamma /{\Gamma_0}){\sigma_0}$ where $\mathfrak{D}$ is the Drude weight and $\Gamma$ is carrier scattering rate and the subscript 0 stands pump off condition. The contribution from Drude weight dominates near charge neutral point and is positive. For higher doping, the contribution from change of scattering rate $\Delta \sigma_{\Gamma} = - (\Delta\Gamma / \Gamma_0) \sigma_0$  controls the $\Delta\sigma$. The authors assumed
\cite{Shi2014} $\Delta\Gamma / \Gamma_0  = 0.2$, independent of $E_F$ to explain the negative dynamic conductivity. This framework does not explain the observed saturation behavior of the dynamic conductivity at higher Fermi energy. Therefore, the issue of the optical pump induced terahertz conductivity, in particular its sign and amplitude is still open and needs to be understood quantitatively.

In this work, we report the THz conductivity $\sigma (\omega)$ of (1) as-grown single layer graphene (AG) by chemical vapor deposition which is unintentionally hole doped, (2) nitrogen doped monolayer graphene (NDG) and (3) thermally annealed nitrogen doped graphene (TAG). The samples have different values of carrier momentum relaxation time $\tau$ and the Fermi energy $E_F$. The conductivities obtained from the THz transmission measurements compare well with the estimates from the relative Raman intensities of D and G bands. Next, we present optical pump (1.58 eV) - THz probe measurements of the transient THz photoconductive response of the graphene samples. On photo-excitation, the dynamic conductivity ($\Delta\sigma = \sigma \text{(pump on)} - \sigma \text{(pump off)})$ is negative for the AG but positive for the NDG.  In both cases, the \textbar$\Delta\sigma_{max}$\textbar  is $\sim$ 1.5${G_0}$ where $G_0 = 2e^2 /h (= 77.3 \mu S)$ is quantum of conductance. We show that on thermal annealing of the NDG, the $\Delta\sigma$ is once again negative. A quantitative analysis of $\Delta\sigma$ is done by noting that in the terahertz range, intraband scattering contribution to $\Delta\sigma$ is orders of magnitude larger than the interband contribution. We invoke secondary hot carrier generation (SHCG)
\cite{Tielrooij2013} to explain quantitatively the negative $\Delta\sigma$ in AG. The sign and magnitude of the dynamic conductivity in graphene thus depends on the relative contributions of the intraband scattering and the SHCG, which, in turn, depend on the momentum relaxation time and the Fermi energy. The cooling of hot carriers is quantitatively analyzed in terms of super-collision (SC) processes. The frequency dependence of dynamic conductivity is also measured.
 
\section{method}
\subsection{Terahertz Set Up}
The output beam of the Ti: sapphire regenerative amplifier laser system which produces $\sim$50 fs optical pulses at a central wavelength of 785 nm with a repetition rate of 1 KHz is divided into three parts: for terahertz generation, terahertz detection and  optical pumping. The terahertz radiation is generated by co-focusing the fundamental and its second harmonic beam using 10 cm focal length lens to produce plasma in air. The unwanted light following the plasma was blocked using high resistive silicon wafer. The terahertz was collimated and focused on the sample by a pair of off-axis parabolic mirrors and again collimated and focused by parabolic mirrors on a 1 mm thick ZnTe crystal used as detector. All experiments were carried out in nitrogen environment at room temperature. The estimated terahertz electric field is $\sim$30 kV/cm. The optical pump-induced changes in the terahertz transmission were measured at the maximum position of the terahertz electric field. The spot size of the pump beam was $\sim$ 0.7 cm and the spot size of the terahertz beam was $\sim$ 0.3 cm so that the pump can excite the sample uniformly. To measure the photoinduced transmission the chopper (341 Hz) was placed in the pump path and to measure the THz electric field from the unexcited sample, the chopper was placed in the path of the terahertz beam. All experiments were performed in transmission geometry.
\subsection{Sample Preparation}
The graphene samples were grown by using the well-known chemical vapor deposition (CVD) method on 25 $\mu m$ thick copper foil.\cite{Li2009} Before growth, the copper foils were cleaned through a chemical process using acetone, acetic acid, deionised water, isopropyl alcohol and methanol successively. In order to remove the oxides and chemical residues, the copper substrates were heated at 1000$^o$C for 30 minutes in presence of hydrogen at a pressure of 20 Torr. Subsequently, methane was flown into the chamber, and graphene growth was carried out for 30 minute keeping hydrogen and methane at a fixed ratio of 1:3, followed by stoppage of the methane flow and the system allowed to cool at a rate of 20$^o$C/min for first 20 minutes without changing the hydrogen flow rate. Then, the system was cooled by a normal fan up to a temperature of 350$^o$C in 1 hr., followed by natural cooling to room temperature. For nitrogen doping, protocol was similar to that reported recently.\cite{Ruitao2012} Namely, after stopping the methane flow and cooling the system to 850$^o$C, ammonia (10 sccm) was passed for 10 min. Subsequently, ammonia flow was stopped and the system was cooled in hydrogen atmosphere as described before. Graphene was transferred onto the 1 mm thick $\alpha$-quartz by the PMMA technique.\cite{Suk2011}

\section{results and discussions}
\subsection{Characterization of the Sample}
Primarily, the experiments were carried out on two samples: AG and NDG. Later, to check the effect of temperature annealing the NDG was annealed at $400^o$C for 4 hrs in argon atmosphere and all the experiments were also performed on this thermally annealed graphene (TAG). The x-ray photoemission spectroscopy (XPS) was done to confirm the presence and nature of nitrogen in graphene. The samples were further characterized by Raman spectroscopy at room temperature using $\lambda$ = 514 nm of laser light.

\begin{figure}[ht!]
\centering
\includegraphics[width=86mm]{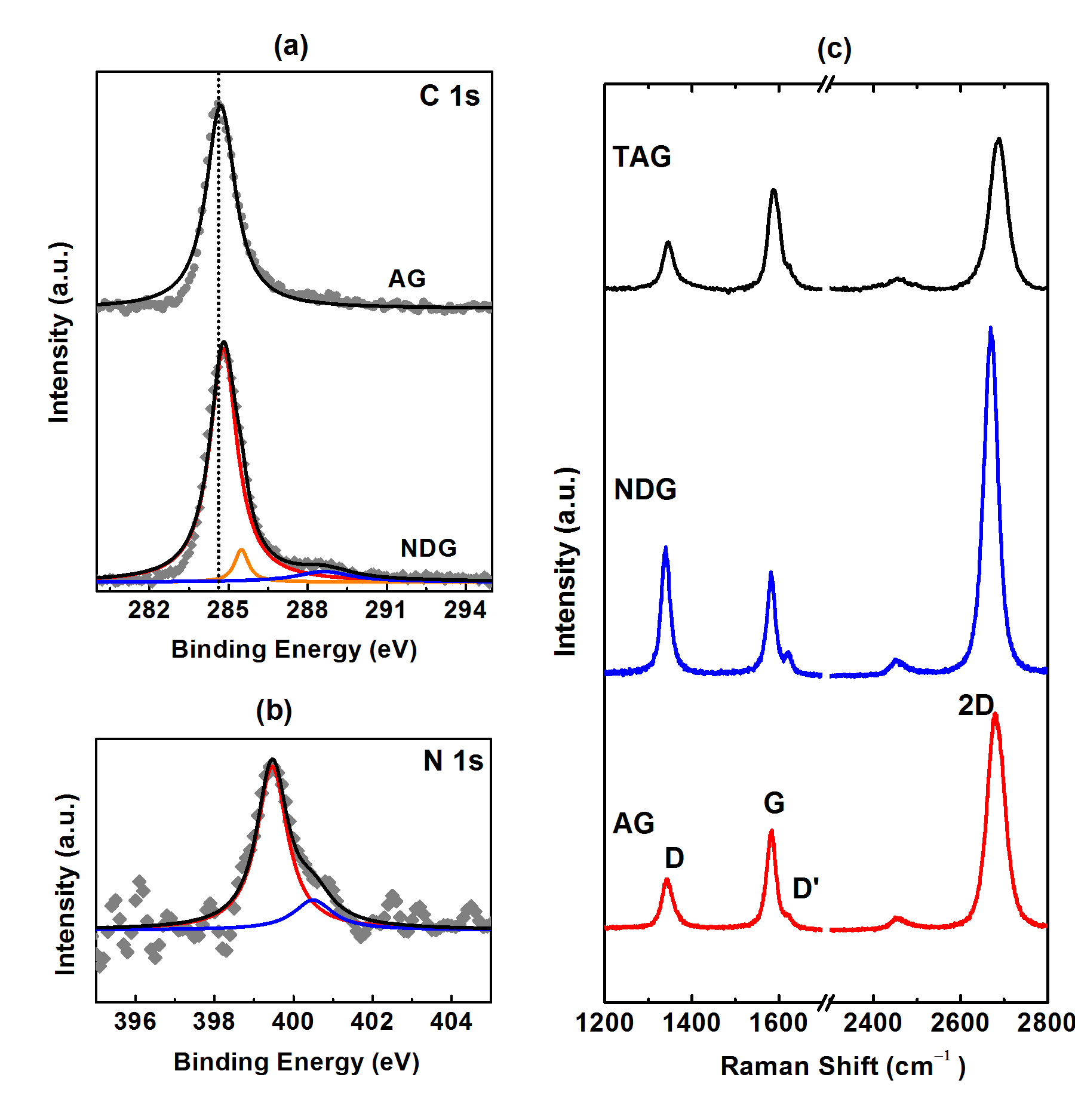}
\caption{(a) XPS spectra of AG and NDG showing the C 1s line. The main peak at 284.6 eV represents $sp^2$ C. Two additional peaks appeared at 285.8 eV and 288.6 eV corresponding to C$-$N and C=O bonds. (b) The N 1s XPS spectra of NDG showing two peak at ~ 399.5 eV and ~ 400.5 eV, which suggest the presence of non-graphitic N $-$ C bonding. (c) Raman spectra of AG, NDG and TAG.}
\label{XPS}
\end{figure}

The C 1s XPS spectra for AG and NDG are shown in Fig.\ref{XPS}a. The peak occurs at binding energy of 284.6 eV for AG corresponding to $sp^2$ carbon.
\cite{Ying2010} For NDG the peaks are at 284.8 eV, 285.8 eV and 288.5 eV corresponding to $sp^2$ carbon, C$-$N and C=O bonds.
\cite{Ying2010} The N 1s XPS spectra of NDG sample shown in Fig.\ref{XPS}b is fitted with two peaks corresponding to non-graphitic substitution namely, pyridinic ($\sim$399.5 eV) and pyrrolic ($\sim$400.5 eV) N.
\cite{Ying2010,Gao2012} In the pyridinic and pyrrolic structures N-atom bonds with two carbon atoms. We cannot resolve any quaternary (graphitic) substitution of N (generally occurs at a binding energy of $\sim$401 eV).
\begin{table*}[ht!] 
\caption{Raman Characterization} 
\centering 
\begin{tabular}{c c c c c c c c c} 
\hline\hline 
Sample & G($cm^{-1}$) & D($cm^{-1}$) & I(2D)/I(G) & A(D)/A(G) & I(D)/I(D$'$) & $E_F$(meV) & $L_a$(nm) & $\tau$(fs)\\ [1ex] 
\hline 
AG & 1583 & 1343 & 2.4 & 0.6 &	7.0 & 180 & 34 & 34 \\ 
NDG & 1582 & 1340	 & 3.5 & 1.3 & 8.7 & 10 & 15 & 15 \\ 
TAG & 1589 & 1346 & 1.5 & 0.5 & 5.7 & 200 & 38 & 38\\ [1ex] 
\hline 
\end{tabular} 
\label{table} 
\end{table*}
Raman spectra of the samples are recorded at room temperature using the excitation wavelength $\lambda$ = 514 nm, displaying four bands (Fig.\ref{XPS}c) where peak positions and relative intensities are given in Table \ref{table}. The G band is symmetry allowed $E_{2g}$ zone-center $\Gamma$ point longitudinal and transverse optical phonon and the D band is associated with the disorder activated near zone-boundary (K point) transverse optical phonon.\cite{Malard2009} The 2D band is due to second-order Raman scattering from near K-point transverse optical phonons and is an unambiguous finger print of the number of layers, as understood by double resonance Raman scattering.
\cite{Malard2009}  The D$'$ band is associated with the disorder activated near $\Gamma$ -point longitudinal optical phonons.The frequencies of the 2D and G bands and their relative intensities depend on the doping levels.
\cite{Das2008} It is known that CVD grown graphene can be unintentionally p-doped due to unwanted charge transfer doping from H$_2$O/O$_2$ molecules.
\cite{Shin2012} Using the intensity ratio of  the 2D and G bands I(2D)/I(G), the Fermi energy of the AG graphene is $E_F$ $\sim$180 meV.\cite{Das2008} After N-doping the intensity ratio I(2D)/I(G) increases which suggests 
\cite{Das2008} that the  N-doping has compensated the p-doping in the AG shifting the Fermi level close to Dirac point with $E_F$ $\sim$10 meV. However the exact value of $E_F$ is difficult to estimate from the Raman data of NDG. The increase in the intensity of the D and D$'$ bands after nitrogen doping clearly reveals an increase in the disorder in NDG. The ratio of the D and G band integrated intensities A(D)/A(G) increase from $\sim$0.6 in AG to $\sim$1.3 in NDG. The ratio A(D)/A(G) is empirically related to the in-plane crystalline grain size without defects ($L_a$) as :
\cite{Cancando2006}   A(D)/A(G) = $[2.4 \times 10^{-10} \text{nm}^3]\lambda^4\text{L}_a^{-1}$ which gives L$_a$ $\sim$34 nm for the AG, $\sim$15 nm for the NDG and $\sim$38 nm for the TAG. Assuming the transport mean free path of the carriers $\ell \sim L_a$ and using Fermi velocity $v_F \approx 1 \times 10^6$ m/s,
\cite{Castro2009} the average momentum relaxation time $\tau \sim \ell /v _F$ is $\sim$15 fs in NDG, 34 fs in AG and 38 fs in TAG. One may try to estimate estimate the conductivity of the graphene in the strong scattering limit
\cite{Tielrooij2013} by using $\sigma = G_0 (E_F\tau / \hbar)$, giving $\sigma \sim 6G_0$ for AG, $\sim 1.4G_0$ for NDG and $\sim 9G_0$ for TAG. The intensity ratio of D and D$'$ bands, I(D)/I(D$'$), has been related to the type of defects.
\cite{Eckmann2012}  For the AG I(D)/I(D$'$) $\sim$7 implies vacancy-like defects;  I(D)/I(D$'$) $\sim$8.7 in NDG implies a presence of sp$^3$ defects ( for $sp^3$ defects,  I(D)/I(D$') \sim$13).
\cite{Eckmann2012}

\begin{figure*}[ht!]
\centering
\includegraphics[width=160mm]{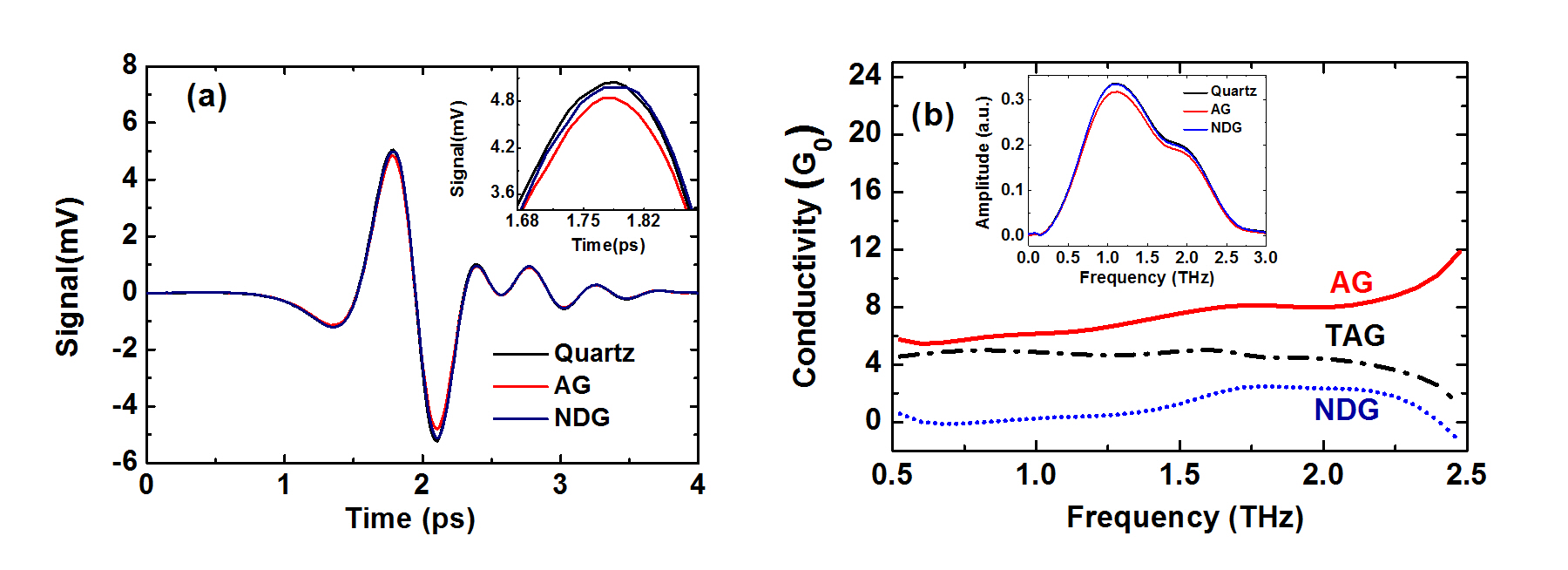}
\caption{Terahertz conductivity. (a) Evolution of THz electric field as a function of delay time. (b) Conductivity (in unit of $G_0$) spectra for AG (red solid line), NDG (blue dotted line) and TAG (black dash-dotted line). Inset: FT amplitudes vs frequency.}
\label{TDspectroscopy}
\end{figure*}
\subsection{Terahertz conductivity of graphene}
Let the temporal evolution of the transmitted terahertz electric fields through the graphene on the $\alpha$-quartz substrate and through the substrate without graphene be denoted by T$_{sam}$(t) and T$_{ref}$(t), respectively. The ratio of Fourier transform (FT) of T$_{sam}$(t) and T$_{ref}$(t) gives amplitude and phase of the spectral transmission function:  S($\omega$)=T$_{sam}$(t)/ T$_{ref}$(t). The complex conductivity spectra can be obtained from the spectral transmission function by using the relation $\text{S}(\omega) = \frac{\text{n}_s +1}{\text{n}_s+1+Z_0\sigma}$  in the limit of thin film approximation.
\cite{Tomaino2011,Tinkham1956}  Here Z$_0$ = 377 $\Omega$ is the impedance of free space and $\text{n}_s$ is refractive index of substrate taken as 2.2. Figure \ref{TDspectroscopy}a shows the temporal terahertz fields through quartz (black line), AG (red line) and NDG (blue line). Figure \ref{TDspectroscopy}b shows the real part of the conductivity in the spectral range 0.5 THz to 2.5 THz and the corresponding FTs are shown in the inset. A nearly spectrally flat conductivity suggests large momentum scattering rate for the graphene samples i.e. $\omega\tau \ll 1$,
\cite{George2008,Strait2011,Liu2011} as also suggested by the estimates of $\tau$ from the Raman data (see Table.\ref{table}). The average conductivities of AG, NDG and TAG samples are (7 $\pm$ 1)G$_0$, (1.5 $\pm$ 0.5)G$_0$ and (4 $\pm$ 1)G$_0$ respectively, in close agreement with the estimates obtained from the Raman data. As mentioned earlier, the reduction of conductivity of the NDG is due to the shift of the Fermi level towards the Dirac point and a decrease of the momentum relaxation time due to increase in disorder.
\begin{figure}[ht!]
\centering
\includegraphics[width=50mm]{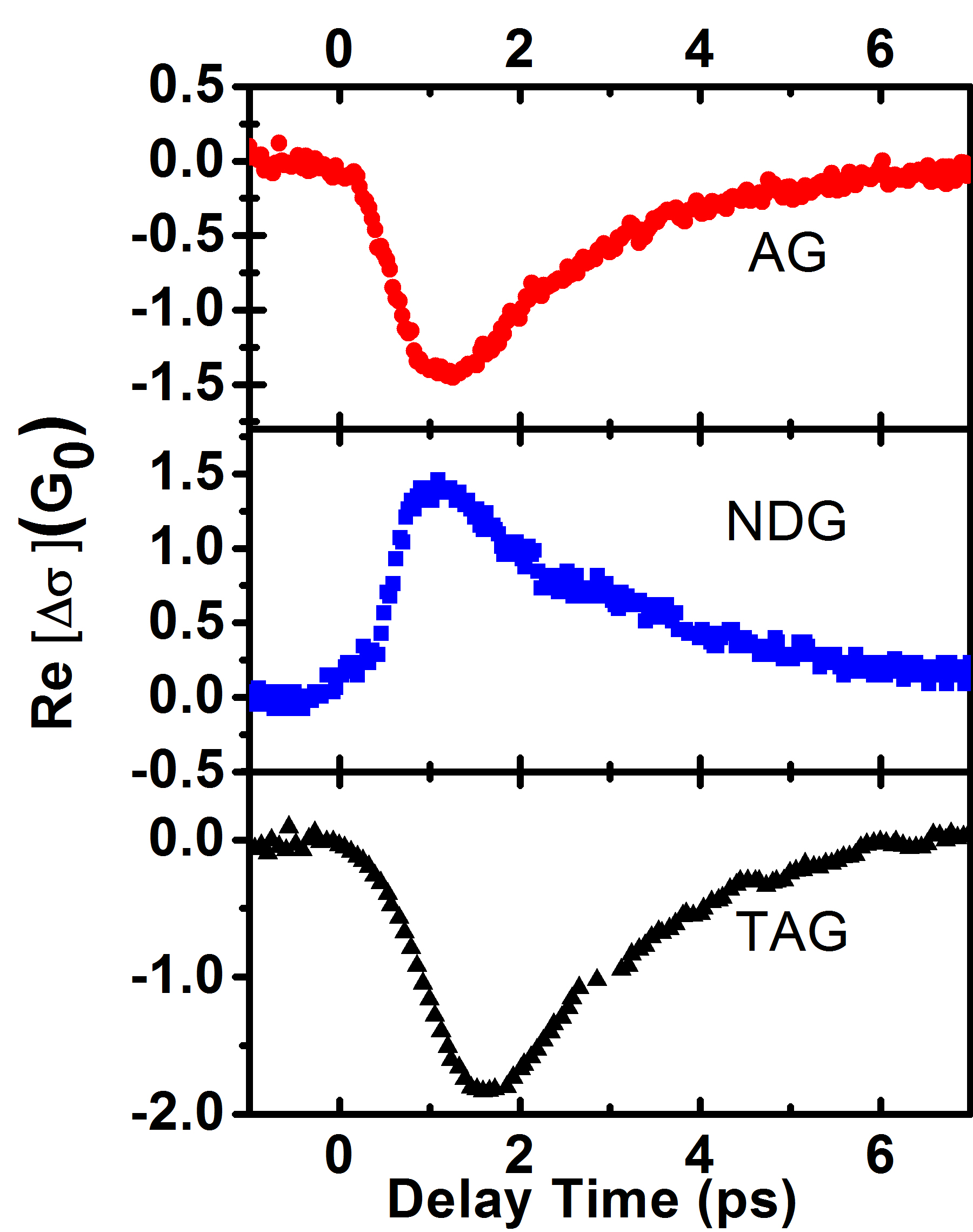}
\caption{Transient THz photoconductivity. Evolution of the real part of photoinduced THz conductivity as function of delay time between 785 nm pump and THz probe. }
\label{OPTP}
\end{figure}
\subsection{Optical Pump Induced Changes in Terahertz Conductivities}
The differential transmission, $\Delta$T/T is related to the dynamic THz conductivity, $\Delta\sigma$ by the relation $\frac {\Delta T}{T} \approx - \frac{Z_0}{n_s + 1}\Delta\sigma$. The results shown in Fig.\ref{OPTP} correspond to the peak of THz electric field using pump excitation density of 340 $\mu$J/cm$^2$ per pulse. Taking the absorption at 785 nm to be 2.3$\%$, photo-excited carrier density is 3 $\times$ 10$^{13}$ /cm${^2}$.
\cite{Mak2008} Most interestingly, $\Delta\sigma$ is negative for the AG whereas it is positive for the NDG. After annealing the NDG, $\Delta\sigma$ is once again negative for TAG.
\begin{figure*}[ht!]
\centering
\includegraphics[width=160mm]{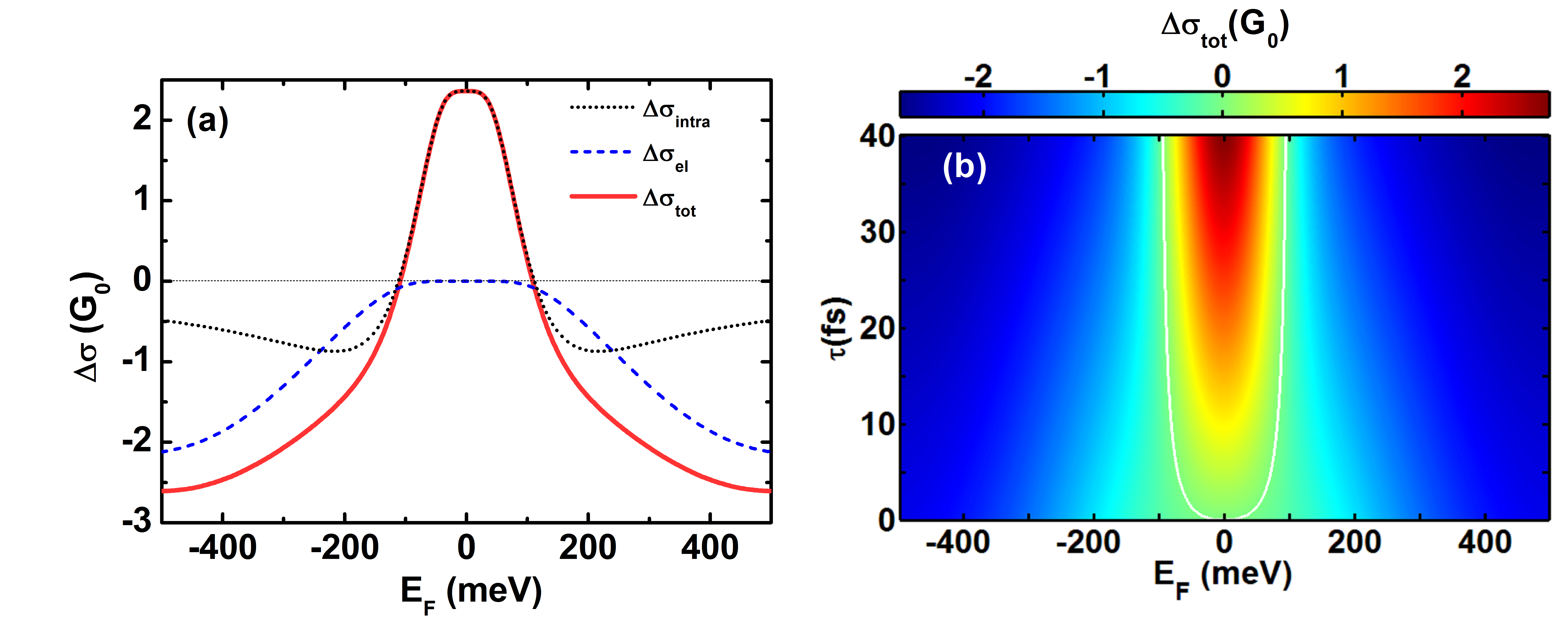}
\caption{scattering mechanisms. (a)$\Delta\sigma_{intra}$ , $\Delta\sigma_{el}$ ,$\Delta\sigma_{tot}$ are plotted as a function of Fermi energy,$ E_F$ at $T_e$ = 700 K, $\tau$ = 34 fs and b = 190 fs/eV. It is positive whatever is the Fermi energy and $T_e$. (b) The contour plot of $\Delta\sigma_{tot}$ as function of Fermi energy and momentum relaxation time $\tau$ at $T_e$ = 700 K and b = 190 fs/eV. The white line represents $\Delta\sigma_{tot}$ = 0 and indicates the boundary between positive and negative $\Delta\sigma_{tot}$. }
\label{model}
\end{figure*}
We now analyze various contributions to the dynamic conductivity. As mentioned earlier, hot carriers achieve quasi equilibrium Fermi distribution with electron temperature T$_e$, thus $\Delta\sigma = \sigma (T_e) - \sigma (T_0)$, where T$_0$ is the lattice temperature ( = 300 K). The intraband and interband contributions to dynamic conductivity is calculated using.
\cite{Winnerl2011}
\begin{equation}
\sigma_{intra} (T) = \frac{2 G_0 \tau k_B T}{\hbar(\omega^2\tau^2 + 1)} [2\cosh (- \frac{E_F}{2k_B T})]
\label{intra}
\end{equation}
\begin{equation}
\sigma_{inter} (T) = \frac{\pi G_0}{8} [\tanh (\frac{\hbar\omega+2E_F}{4k_B T})+\tanh (\frac{\hbar\omega-2E_F}{4k_B T})]
\label{inter}	
\end{equation}
where $\omega$ is terahertz probe frequency. The interband contribution
\cite{supporting} 
gives positive and negative $\Delta\sigma$ depending on $E_F$ but is $\sim10^3$ times smaller as compared to the intraband contribution. Hence the photoinduced terahertz conductivity is dominated by the intraband scattering
\cite{George2008,Strait2011,Winnerl2011} and therefore we will neglect interband contributions. Taking $\tau$ = 34 fs, the intraband contribution to the dynamic conductivity at 1 THz is shown in Fig.\ref{model}a (dotted black line) as a function of Fermi energy at a representative T$_e$ = 700 K. The magnitude of $\Delta\sigma_{intra}$ as seen in Fig.\ref{model}a is not sufficient to explain the observed $\Delta\sigma$ in AG (E$_F \sim$180 meV) which necessarily requires another mechanism. We consider the recently proposed secondary hot carrier generation (SHCG)
\cite{Tielrooij2013} in which the photoexcited carriers interact with the intrinsic carriers to excite the later. The photoexcited carriers have two scattering channels; one is the conventional intraband scattering mechanism with momentum relaxation time, $\tau$ as discussed earlier and the other is the Coulomb scattering with momentum relaxation time, $\tau_e$ which is proportional to carrier energy $\varepsilon$ i.e.,$\tau_e = b\varepsilon$ where b, the proportionality constant, depends on the ratio of the average inter-electron Coulomb interaction energy to the Fermi energy and the density of the secondary hot carriers (n$_i$) (see Eq. 3.21 of Ref.\onlinecite{Das2011}). The secondary hot carrier generation contribution to the photoinduced conductivity is given by
\cite{Tielrooij2013}

\begin{equation}
\Delta\sigma_{el}=k_B^2(T_e^2 - T_0^2)\frac{\pi^2}{6}\nu(E_F)\frac{\partial ^2F(\varepsilon)}{\partial\varepsilon^2}\vert_{\varepsilon=E_F},
\label{SHCG}
\end{equation}
\begin{equation}
F (\varepsilon)=e^2 v_F^2 \frac{\tau_e(\varepsilon)}{1+\omega^2[\tau_e(\varepsilon)]^2}
\end{equation}
where $\nu(E_F)$ is density of states of graphene at Fermi energy. As suggested in Ref .\onlinecite{Tielrooij2013} we take $\tau_e$ = 34 fs and E$_F$ = 180 meV to get b = 190 fs/eV in calculating $\Delta\sigma_{el}$. Figure \ref{model}a (dashed blue line) shows that $\Delta\sigma_{el} < 0$ and the magnitude is comparable to the intraband contribution to the terahertz dynamic conductivity. The total change of terahertz photoconductivity is the sum of intraband and the SHCG contributions: $\Delta\sigma_{tot}=\Delta\sigma_{intra}+\Delta\sigma_{inter}$. Figure \ref{model}b shows the dynamics terahertz photoconductivity at 1 THz as a function of Fermi energy $E_F$ and momentum relaxation time $\tau$ at a representative electron temperature of 700 K. It shows that the magnitude and sign of $\Delta\sigma_{tot}$ is determined by the value of the Fermi energy and the momentum relaxation time. In AG, $E_F \sim$180 meV and $\tau \sim$34 fs, resulting in negative $\Delta\sigma_{tot}$. In contrast, a shift of Fermi level towards the Dirac point in NDG results in $\Delta\sigma_{tot} > 0$ due to dominance of the intraband scattering process. Again, in TAG the increased Fermi energy lead to $\Delta\sigma_{tot} < 0$. The results are similar for other representative values of parameter b (b = 120, 300 fs/ev) \cite{supporting} 
\begin{figure*}[ht!]
\centering
\includegraphics[width=160mm]{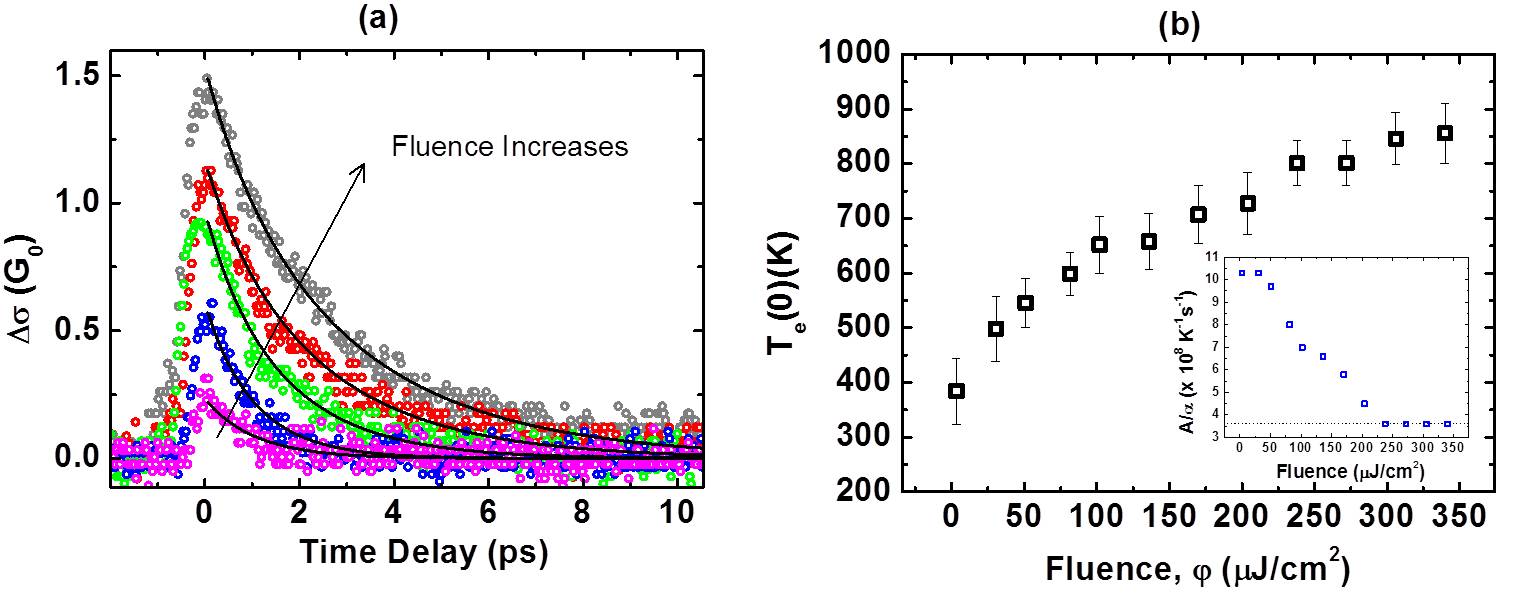}
\caption{Cooling dynamics of NDG. (a) OPTP experimental data for fluences 340, 238, 102, 30 and 3 $\mu \text{J/cm}^{2}$ are shown by opened circles of color gray, red, green, blue and magenta respectively and solid lines are the fittings to the SC cooling mechanism. (b) Quasi-equilibrium electron temperature ($T_e$) as a function of fluence. Inset: the fitted parameter A/$\alpha$ is plotted as a function of fluence. }
\label{NDGfit}
\end{figure*}

\begin{figure*}[ht!]
\centering
\includegraphics[width=160mm]{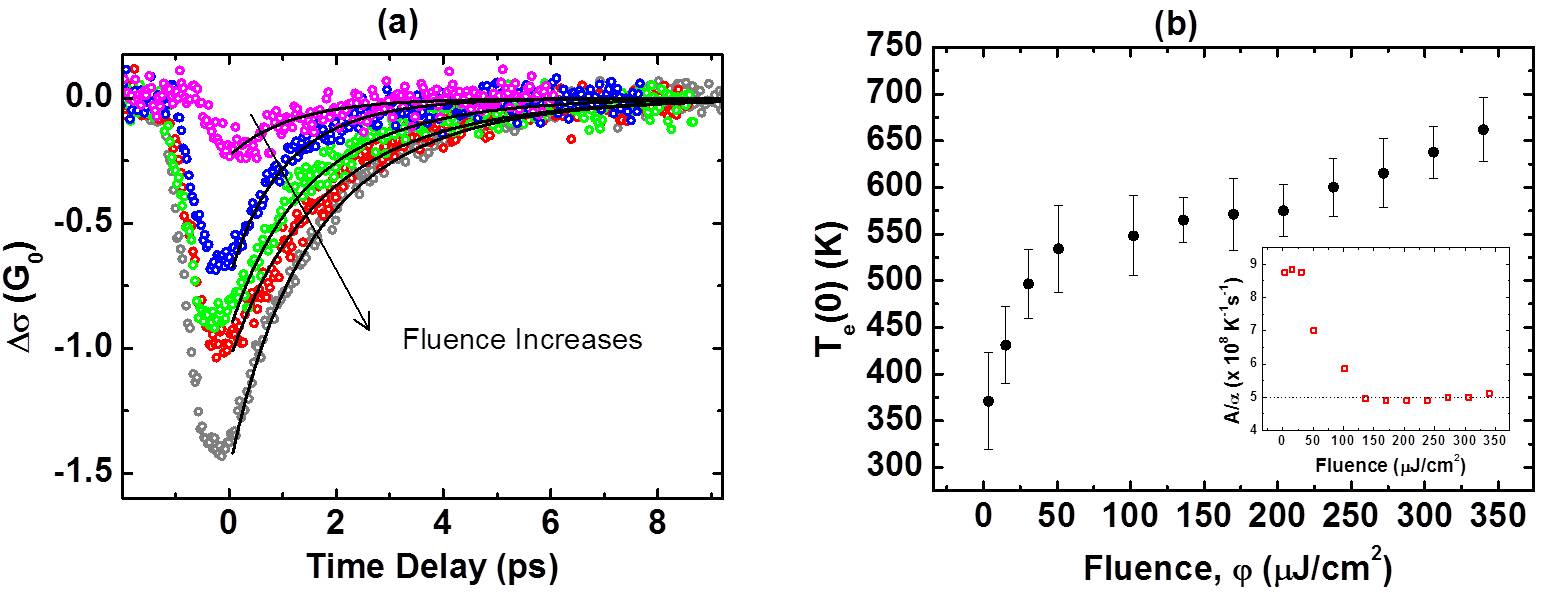}
\caption{Cooling dynamics of AG. (a) OPTP experimental data for fluences 340, 238, 102, 30 and 3 $\mu \text{J/cm}^{2}$  are shown by opened circles of color gray, red, green, blue and magenta respectively and solid lines are the fittings to the SC cooling mechanism. (b) Quasi-equilibrium electron temperature ($T_e$) as a function of fluence. Inset: the fitted parameter A/$\alpha$ is plotted as a function of fluence.  }
\label{AGfit}
\end{figure*}

\subsection{Cooling Dynamics}
We now focus on the cooling dynamics of the hot carriers. It is known
\cite{Gierz2013,Lui2010,Kampfrath2005} that photoexcited electrons can efficiently lose their energy by emitting optical phonons with energy $\sim$200 meV with a relaxation time of $\sim$300 to 500 fs. Below 200 meV, the hot carriers can dissipate energy only by emitting acoustic phonons with energy $k_BT_{BG}$ per scattering event as permitted by conservation of momentum. Here, $T_{BG}$ is Bloch-Grüneisen temperature given by $k_BT_{BG}=(2v_s/v_F)E_F$ ($v_s \sim 2.1 \times 10^4 m/s$ is velocity of sound in graphene) which defines a boundary between the low and the high temperature behavior. The cooling of carriers can last as long as 300 ps for phonon temperatures $T_0 > T_{BG}$. However, according to the super collision (SC) model\cite{Song2012109} an alternate route of energy relaxation of the hot carriers is via the disorder-mediated emission of high energy ($\sim k_BT_e$) and high momentum ($\sim k_BT_e/ \hbar v_s$) acoustic phonons. The enhancement factor for energy dissipation rate over momentum conserving path ways is expressed as
\cite{Song2012109}
\begin{equation}
\frac{H_{SC}}{H_0}=\frac{0.77}{k_F \ell}\frac{T_e^2+T_eT_0+T_0^2}{T_{BG}^2}
\label{ratio}
\end{equation}
Therefore, the enhancement factor depends on temperature, Fermi energy and disorder. The temperature $T_{BG}$ is $\sim$87 K for AG and $\sim$5 K for NDG which are less than $T_0$(= 300 K). According to Eq. \ref{ratio} the enhancement factor is $\sim$17380 for NDG and $\sim$13 for AG, taking $T_e$ = 700 K and $T_0$ = 300 K showing the dominance of the SC cooling.

For the SC mechanism the heat dissipation rate is given by\cite{Song2012109}  $H_{SC}=A(T_e^3-T_0^3)$ where  $A=9.62\frac{gD_{ac}E_Fk_B^3}{\hbar k_F \ell}$ where g is electron-phonon coupling strength and $D_{ac}$ is the deformation potential. After acheiving the quasi-equilibrium temperature $T_e(0)$, subsequent decreases of the carrier temperature  due to the SC is described by
\cite{Song2012109}
\begin{equation}
\frac {\partial T_e(t)}{\partial t}=-\frac{H_{SC}}{\alpha T_e}=-\frac{A}{\alpha} \frac {T_e^3-T_0^3}{T_e(t)}
\label{rateEQ}
\end{equation}
 where $A/ \alpha$ is the SC rate coefficient. To fit the data shown in Fig.\ref{NDGfit}a, one should in principle use a sum of two terms: first term is an exponential term with relaxation time hundreds of femtosecond to represent the contribution of the optical phonons and a second term given by the solution of Eq.\ref{rateEQ}  for the SC mechanism. However we find that our data is not very sensitive to the contributions from the  term representing the optical phonons and hence  Figure \ref{NDGfit}a shows the fitting of transient conductivity of the NDG with the SC model described by Eq.\ref{rateEQ} with $A/ \alpha$ and $T_e(0)$ as fitting parameters. Here, we have used Eq.\ref{intra} and Eq.\ref{SHCG} to calculate dynamic conductivity, taking $\tau$ = 15 fs and $E_F$ = 10 meV as suggested by our Raman data. Figure \ref{NDGfit}b  shows fluence dependence of the fitting parameter $T_e(0)$ and the inset shows saturation of $A/ \alpha$ at high fluence suggesting that the  SC dominates over momentum conserving paths at higher fluences.

The SC cooling rate coefficient is :\cite{Song2012109}
 \begin{equation}
\frac {A}{\alpha}=\frac {6\zeta(3)}{\pi^2}\frac {g}{k_F\ell}\frac{k_B}{\hbar}\cong\frac{2}{3}\frac {g}{k_F\ell}\frac{k_B}{\hbar},
\label{ratecoefficient}
 \end{equation}
where
\begin{equation}
g=\frac{D_{ac}^2}{\rho v_s^2}\frac{2E_F}{\pi(\hbar v_F)^2}
\end{equation}
where $\zeta$ is Reimann zeta function and $\rho\sim7.6 \times 10^{-7}$ kg/m$^2$  is density of graphene. Therefore, the SC rate coefficient depends on the graphene sample and will not vary with the number of photoexcited carriers at higher fluence where the SC cooling dominates. The SC cooling dominated $A/ \alpha$ can be used to estimate the deformation potential of graphene. Taking $E_F$ = 10 meV, $k_F\ell = \sigma (h/2e^2) \sim1.4 $ and $A/ \alpha = 3.6 \times 10^8 K^{-1}s^{-1}$, the deformation potential is $D_{ac} \sim$28 eV, comparable to the reported values of $D_{ac}\sim$ 10 - 30 eV.
\cite{Bolotin2008,Dean2010}

In a similar manner, the relaxation dynamics of AG is fitted with SC model using Eq. \ref{intra}, \ref{SHCG} and \ref{rateEQ}. The fitted graphs shown in Fig.\ref{AGfit}a are generated using $\tau$ = 34 fs, b = 190 fs/eV and $E_F$ (at 300 K) = 180 meV. The fluence dependence of the fitted parameters $T_e(0)$ and $A/ \alpha $ are shown in Fig.\ref{AGfit}b. At higher fluences $A/ \alpha$ saturates at $\sim5.0 \times 10^8 K^{-1}s^{-1}$ (shown in the inset), which gives (using $k_F\ell \sim$6) the deformation potential $D_{ac} \sim$ 17 eV. Since, the electron heat capacity of graphene is proportional to density of state and hence the Fermi energy,
\cite{Shi2014, Song2012109} the NDG will have smaller electronic heat capacity than that of AG and therefore the electron temperature $T_e$ in NDG will be more than that in AG for a given pump fluence. This is indeed the case since $T_e(0) = (680 \pm 34)$ K in AG and $T_e(0) = (855 \pm 50)$ K in NDG at $\varphi = 340 ~\mu J/cm^2$.
\subsection{Frequency dependence of dynamic conductivity}
\begin{figure*}[ht!]
\centering
\includegraphics[width=178mm]{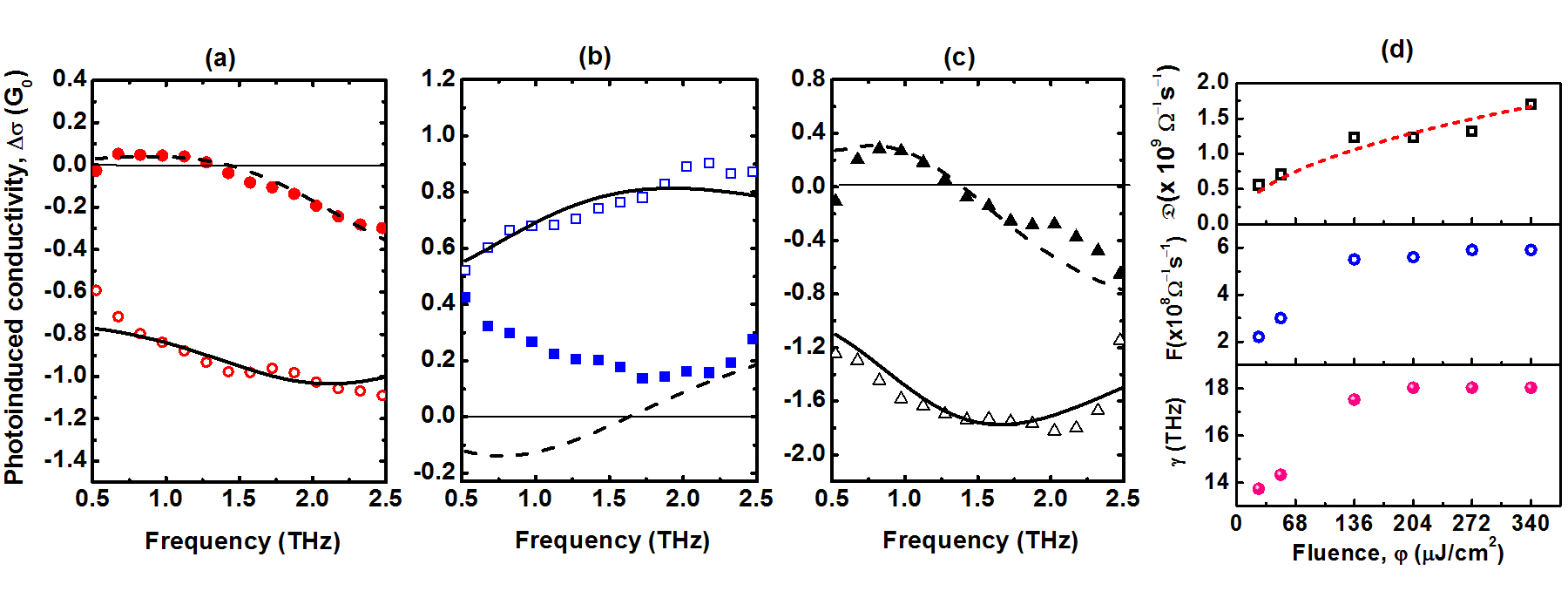}
\caption{Photoinduced conductivity $\Delta\sigma(\omega)$ as a function of frequency at 2 ps after photoexcitation for (a) AG, (b) NDG and (c) TAG. The real parts of the data are shown by opened symbol and imaginary parts are shown by closed symbols. The continuous horizontal lines shows the zero reference lines of $\Delta\sigma(\omega)$. The black and dashed nonlinear black lines are the fit to the Drude-Lorentz model (Eq. \ref{DrudeLorentz}).  (d) Fitted parameters for AG. The Drude weight, $\mathfrak{D}$ varies as $\varphi ^{1/2}$ (shown by dotted red line). }
\label{fre}
\end{figure*}
We now present the frequency dependence of the dynamic conductivity. After the pump excitation, the photoinduced change in terahertz electric field, $\Delta$T(t) throughout the complete terahertz pulse is measured 
\cite{supporting}
for a pump fluence of $\sim340 \mu J/cm^2$ and Figure \ref{fre}a, \ref{fre}b and \ref{fre}c show the corresponding real and imaginary part of photoinduced conductivity for AG, NDG and TAG respectively. In AG the zero crossing of the imaginary part of $\Delta\sigma$ at 1.3 THz strongly suggests a corresponding peak in the real part of $\Delta\sigma$ which can be described by a Lorentzian oscillator. However, the amplitude of the imaginary part is much less than the real part and points to the Drude behavior. Hence Drude-Lorentz model has been used to describe the frequency dependence of $\Delta\sigma(\omega)$ 
\cite{Docherty2012,Parkinson2007, Parkinson2009} 
\begin{equation}
\Delta\sigma =\frac {\mathfrak{D}\tau}{1 - i\omega t} + \frac{iF\omega}{(\omega^2-\omega_0^2)+i\omega\gamma}
\label{DrudeLorentz}
\end{equation}
Here, the first term is the Drude part in terms of Drude weight $\mathfrak{D}$ and momentum relaxation time $\tau$. The second term is the Lorentz part with F as the oscillator strength, $\gamma$ as linewidth and $\omega_0$ the resonant frequency. The fitted curves are shown in Fig. \ref{fre}a for the AG ($\mathfrak{D}$ = $1.7 \times 10^9 \Omega^{-1}s^{-1}$, $\tau$ = 34 fs, F = $5.9 \times 10^8 \Omega^{-1}s^{-1}$, $\gamma$ = 18 THz, $\omega_0/2\pi$ = 2.3 THz). In comparison, for the NDG the imaginary part is positive as in Drude-like response but the positive real part shown in Fig. \ref{fre}b cannot be explained by the Drude model. The Drude–Lorentz model (Eq. \ref{DrudeLorentz}) is not able to fit the  $\Delta\sigma(\omega)$ (Fig. \ref{fre}b). A poor fit to the imaginary part of  $\Delta\sigma(\omega)$ needs to be understood. Here, we have used $\tau$ = 15 fs (obtained from Raman scattering) and $\omega_0/2\pi$  = 2.0 THz.  The $\Delta\sigma(\omega)$ of TAG is nearly same as that of the AG and fitted with Drude Lorentz model with $\tau$ = 38 fs and  $\omega_0/2\pi$ = 1.7 THz, as shown in Fig. \ref{fre}c.

The dynamic conductivity $\Delta\sigma(\omega)$ of the AG is plotted at various the pump fluence varying from 25 to $340 ~ \mu J/cm^2$.
\cite{supporting}
The fittings are performed by taking $\tau$ = 34 fs and $\omega_0/2\pi$  = 2.3 THz and varying $\mathfrak{D}$, F and $\gamma$. The fluence dependence of the parameters $\mathfrak{D}$, F and $\gamma$ are shown in Fig. \ref{fre}d. It is seen that $\mathfrak{D}$ $\sim \varphi^{1/2}$ (see the dotted line in top panel) as expected for graphene ($\mathfrak{D} = (v_Fe^2/\hbar)\sqrt{\pi\vert n\vert}$ ). The parameters F and $\gamma$ saturate at fluence above $\sim100 ~ \mu J/cm^2$. The physical significance of the resonant frequency $\omega_0$ in Lorentz part of $\Delta\sigma$ is not clear. Docherty et al.\cite{Docherty2012} have attributed this to the opening of a gap in the density of state.

\section{conclusion}
In summary, we have presented a quantitative framework of THz dynamic conductivity in monolayer graphene. We showed that $\Delta\sigma$ is determined by the relative contributions of the secondary hot carrier generation and conventional intraband scattering which, in turn, depend on the position of the Fermi level and momentum relaxation time. In highly doped sample, the photoexcited hot electrons interact with the intrinsic carriers to generate secondary hot carriers which result in decrease of the THz conductivity. In NDG, the Fermi energy is closer to the Dirac point and the enhanced disorder decreases the momentum relaxation time which makes the intraband scattering as the dominant scattering mechanism resulting in positive $\Delta\sigma$. The cooling dynamics of the hot carriers is well explained in both the sample by the disorder mediated electron-acoustic phonon interaction, giving the deformation potential comparable to the previous studies.

\begin{acknowledgments}
AKS thanks Nano Mission Project under Department of Science and Technology, India for funding. We thank Gyan Prakash for his help in the initial part of the experiments.
\end{acknowledgments}

\bibliography{reference}
\bibliographystyle{apsrev4-1}
\end{document}


\title{Tuning Photoinduced Terahertz Conductivity in Monolayer Graphene: Optical Pump Terahertz Probe Spectroscopy}
\author{Srabani Kar}
\affiliation{Department of Physics, Indian Institute of Science, Bangalore 560 012, India}
\affiliation{Center for Ultrafast Laser Applications, Indian Institute of Science, Bangalore 560 012, India}
\author{Dipti R. Mohapatra}
\affiliation{Department of Physics, Indian Institute of Science, Bangalore 560 012, India}
\author{Eric Freysz}
\affiliation{University of Bordeaux, LOMA, UMR CNRS 5798, 351, Cours de la liberation, 33405 Talence cedex, France}
\author{A. K. Sood}
\email[corresponding author:]{asood@physics.iisc.ernet.in}
\affiliation{Department of Physics, Indian Institute of Science, Bangalore 560 012, India}
\affiliation{Center for Ultrafast Laser Applications, Indian Institute of Science, Bangalore 560 012, India}
\maketitle

\begin{figure*}[ht]
\renewcommand{\thefigure}{s\arabic{figure}}
\centering
\includegraphics[width=100mm]{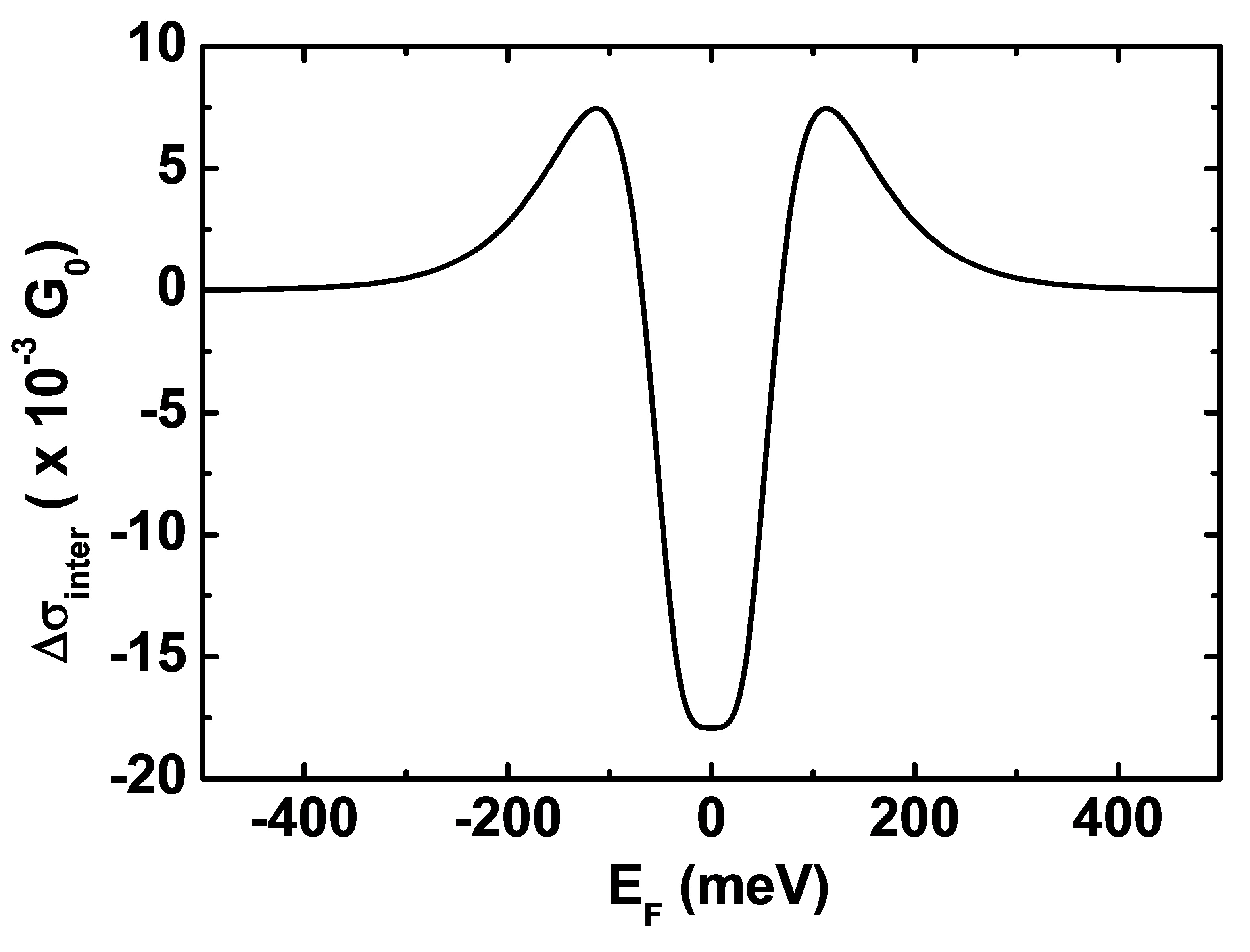}
\caption{The contribution of inter band scattering as a function of Fermi energy at electron temperature $T_e$ = 700 K.}
\label{overflow}
\end{figure*}
\begin{figure*}[ht!]
\renewcommand{\thefigure}{s\arabic{figure}}
\centering
\includegraphics[width=100mm]{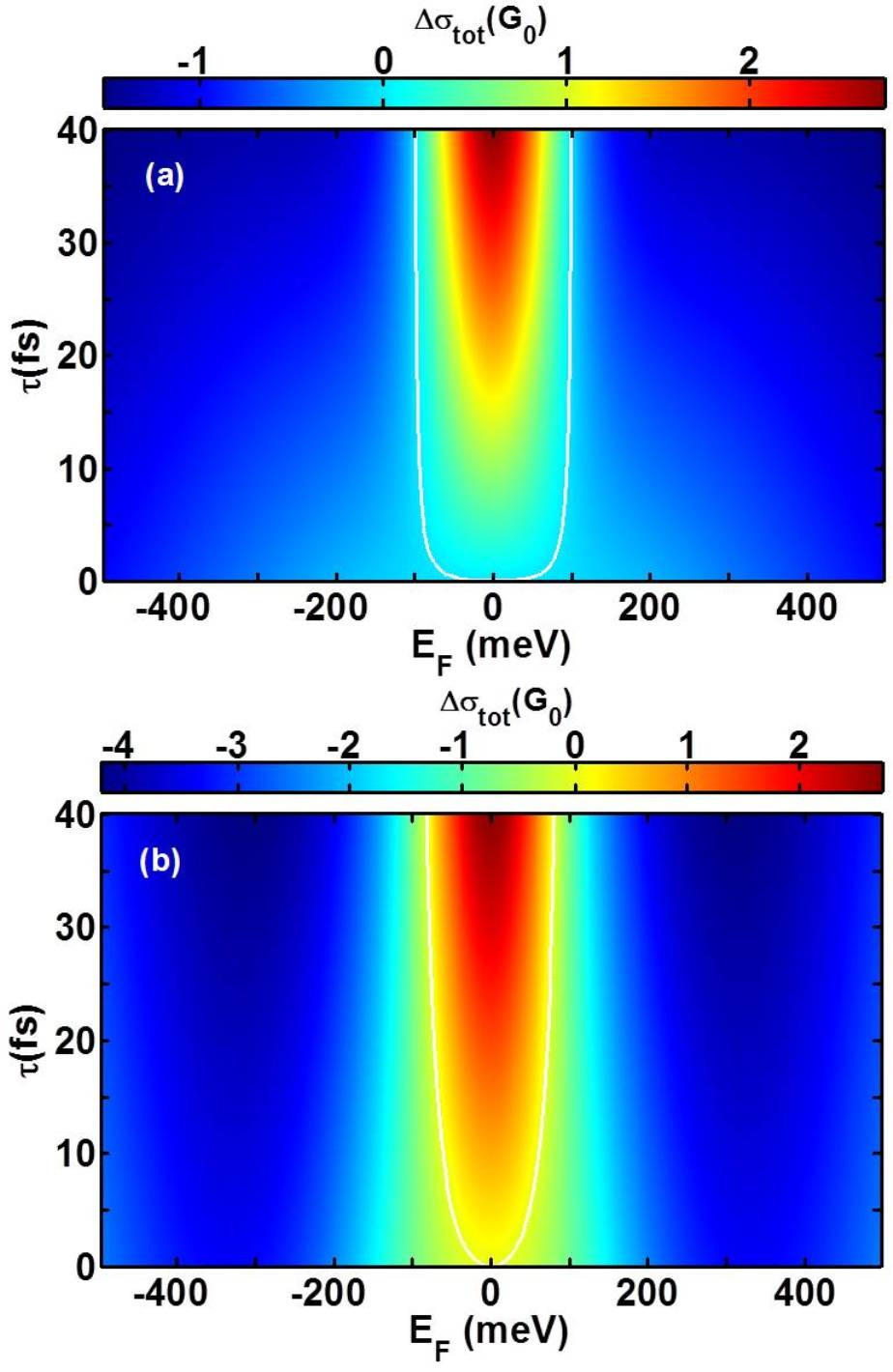}
\caption{Contour plot of $\Delta\sigma_{tot}$  at $T_e$ = 700 K for (a) b = 120 fs/eV and (b) b = 300 fs/eV.}
\label{overflow}
\end{figure*}
\begin{figure*}[ht!]
\renewcommand{\thefigure}{s\arabic{figure}}
\centering
\includegraphics[width=100mm]{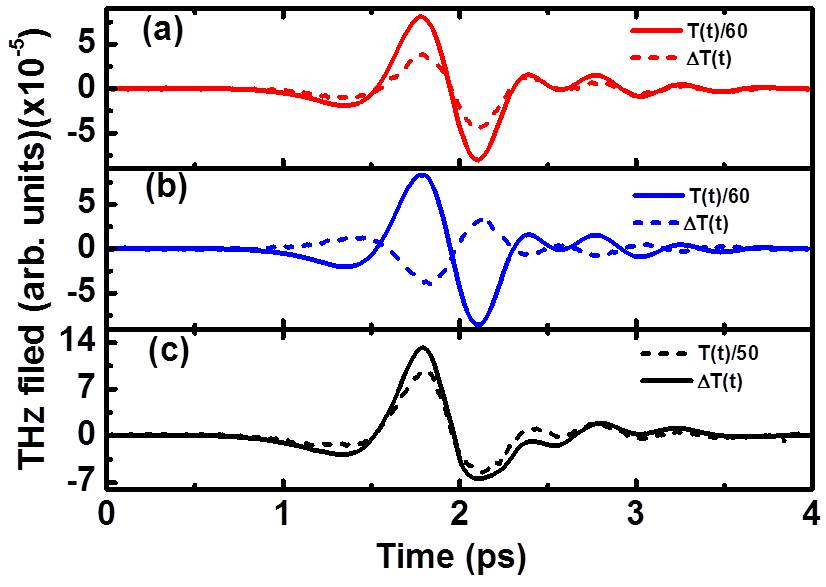}
\caption{Photoinduced change of THz field transmission for (a) as-prepared single layer graphene (AG) (b) doped single layer graphene (NDG) and (c) thermally annealed single layer graphene (TAG) are shown by dashed lines. The Terahertz fields without pump excitation of the corresponding sample are shown by solid lines}
\label{overflow}
\end{figure*}
\begin{figure*}[ht!]
\renewcommand{\thefigure}{s\arabic{figure}}
\centering
\includegraphics[width=100mm]{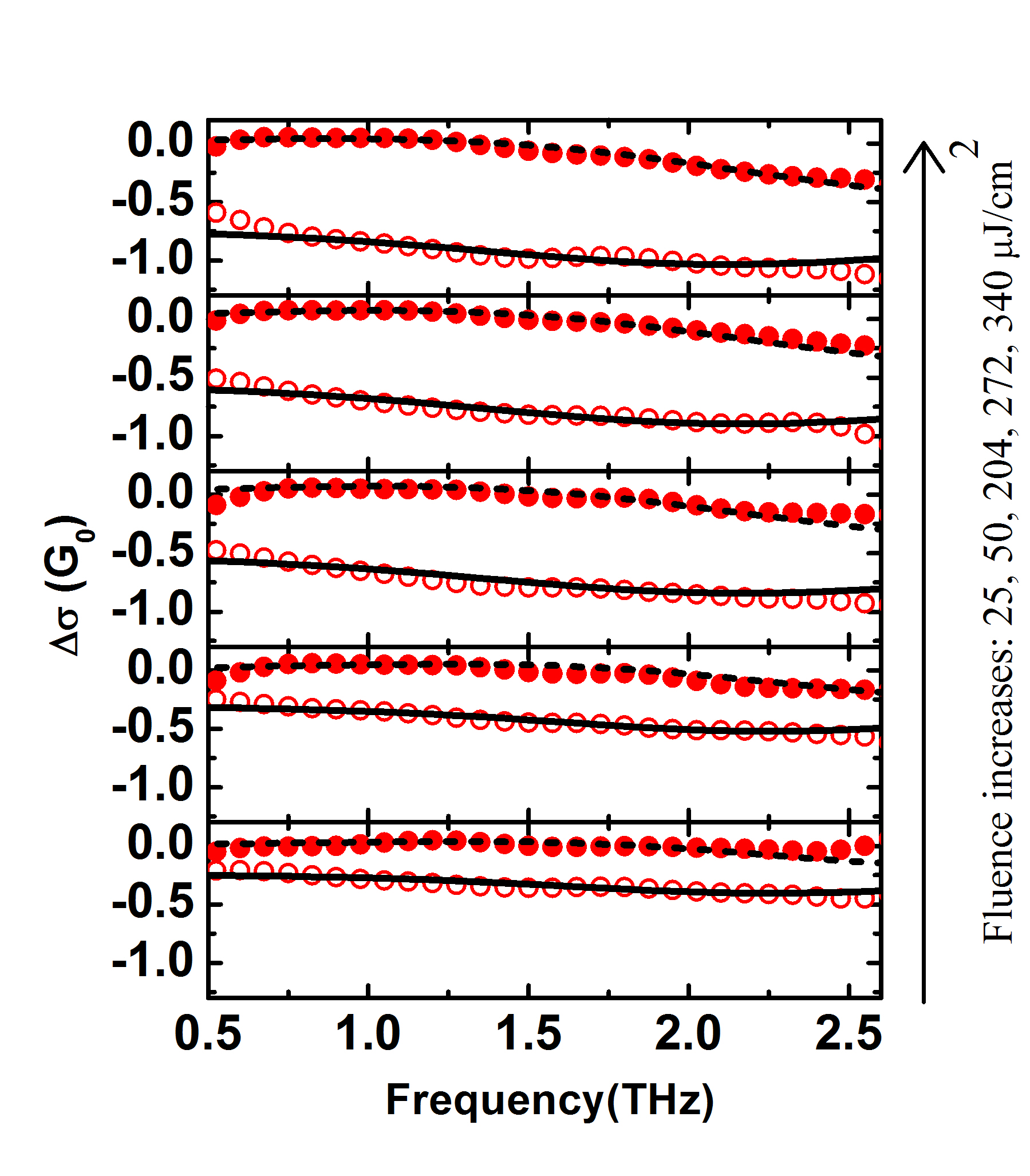}
\caption{The fitted $\Delta\sigma(\omega)$ for pump fluence varying from 25 to 340 $\mu$J/cm$^2$ for AG. The real part is represented by open circles and imaginary part by closed circles.}
\label{overflow}
\end{figure*}